# Exploratory Study of the 3-Gluon Vertex on the Lattice

C. Parrinello*

*Department of Physics, University of Edinburgh,*

*Mayfield Road, Edinburgh EH9 3JZ, U.K.*

(May 26, 1994)

## Abstract

We define and evaluate on the lattice the amputated 3-gluon vertex function in momentum space. We give numerical results for $16^3 \times 40$ and $24^3 \times 40$ quenched lattices at $\beta = 6.0$. A good numerical signal is obtained, at the price of enforcing the gauge-fixing condition with high accuracy. By comparing results from two different lattice volumes, we try to investigate the crucial issue of finite volume effects. We also outline a method for the lattice evaluation of the QCD running coupling constant as defined from the 3-gluon vertex, while being aware that a realistic calculation will require larger $\beta$ values and very high statistics.

*Electronic address: claudio@th.ph.ed.ac.uk

## I. INTRODUCTION

The task of measuring the QCD coupling constant at energies of the order of the $b-$quark mass is a major challenge for the lattice community. Many different methods have been proposed in this respect, based on the study of the interquark static potential, charmonium spectra and other quantites [1–3]. In spite of the success of some of these methods (at least when applied in the quenched approximation), it would be desirable to investigate a more fundamental, field-theoretic definition of the coupling constant, arising from the lattice study of some fundamental QCD interaction vertex, and compare the results with what is obtained from different methods. We discuss here some studies in this direction [4].

On general grounds, the lattice study of gluon (and quark) correlation functions is very interesting because it provides a valuable tool to gain insight into the non-perturbative QCD dynamics at the fundamental level. In fact, in this approach one deals with the fundamental degrees of freedom of the theory, assuming we are sufficiently close to the continuum limit, and the results can be compared in principle to continuum non-perturbative predictions from Schwinger-Dyson equations.

Using a lattice definition of the gluon, one can evaluate any unrenormalized, complete $n$-point Green's function (we omit Lorentz indices):

$$G^{(n)}_{lat}(x_1, \ldots, x_n) \ \equiv < A_{lat}(x_1), \ldots, A_{lat}(x_n) >, \tag{1}$$

or, in momentum space:

$$G^{(n)}_{lat}(p_1, \ldots, p_n) \ \delta^4(p_1 + \ldots + p_n). \tag{2}$$

Once $G^{(2)}_{lat}(p^2)$ and $G^{(3)}_{lat}(p_1, p_2, p_3)$ are determined, many interesting issues can be investigated. The first one is the non-perturbative behaviour of the gluon propagator $G^{(2)}(p^2)$, which has been analyzed by many authors on the lattice [5–9], yet it is still not completely understood. Secondly, from $G^{(2)}_{lat}$ and $G^{(3)}_{lat}$ one may define non-perturbatively the lattice version of the amputated, 1PI 3-gluon vertex function:

$$\Gamma^{(3)}_{lat}(p_1,p_2,p_3) \equiv G^{(3)}_{lat}(p_1,p_2,p_3) \times \prod_{i=1}^{3} \left[G^{(2)}_{lat}(p_i^2)\right]^{-1}. \tag{3}$$

Evaluating the above quantity is the crucial step in order to define a QCD coupling constant from the lattice 3-gluon vertex, as we explain in the following [4].

In order to determine a convenient kinematical setup, we first consider the general form of the continuum, off-shell 3-gluon vertex [10], which can be written as

$$\Gamma^{(3)}_{cont\ \alpha\beta\gamma} = \Gamma^{(3)\ T}_{cont\ \alpha\beta\gamma} + \Gamma^{(3)\ L}_{cont\ \alpha\beta\gamma}, \tag{4}$$

where

$$\begin{aligned}
&\Gamma^{(3)\ T}_{cont\ \alpha\beta\gamma}(p_1,p_2,p_3) = F(p_1^2,p_2^2;p_3^2)(g_{\alpha\beta}\ p_1\cdot p_2 - p_{1\beta}p_{2\alpha})B^3_\gamma \\
&+H\ [-g_{\alpha\beta}\ B^3_\gamma + \frac{1}{3}\ (p_{1\gamma}\ p_{2\alpha}\ p_{3\beta} - p_{1\beta}\ p_{2\gamma}\ p_{3\alpha})] +\ cyclic\ permutations,
\end{aligned} \tag{5}$$

with

$$B^3_\gamma = (p_{1\gamma}\ p_2\cdot p_3 - p_{2\gamma}\ p_1\cdot p_3), \tag{6}$$

and

$$\begin{aligned}
&\Gamma^{(3)\ L}_{cont\ \alpha\beta\gamma}(p_1,p_2,p_3) = \mathcal{A}(p_1^2,p_2^2;p_3^2)\ g_{\alpha\beta}\ (p_1-p_2)_\gamma + B(p_1^2,p_2^2;p_3^2)\ g_{\alpha\beta}\ (p_1+p_2)_\gamma \\
&+C(p_1^2,p_2^2;p_3^2)\ (p_{1\beta}\ p_{2\alpha} - g_{\alpha\beta}\ p_1\cdot p_2)\ (p_1-p_2)_\gamma + \frac{S}{3}\ (p_{1\gamma}\ p_{2\alpha})\ p_{3\beta} + p_{1\beta}\ p_{2\gamma}\ p_{3\alpha}) +\ c.p.
\end{aligned} \tag{7}$$

In the above expressions the scalar functions $F, \mathcal{A}$ and $C$ are symmetric in their first two arguments, $B$ is antisymmetric in the first two arguments, $S$ is antisymmetric under exchange of any pair of arguments and $H$ is totally symmetric in $p_1^2, p_2^2, p_3^2$.

Summarizing, $\Gamma^{(3)}_{cont\ \alpha\beta\gamma}$ contains 6 independent scalar functions, but for the purpose of computing the coupling constant renormalization one only needs to determine the function which multiplies the tree-level vertex, namely $\mathcal{A}(p_1^2,p_2^2;p_3^2)$. Of course this is the only one which is divergent when the UV cutoff is removed.

If one evaluates $\Gamma^{(3)}_{cont\ \alpha\beta\gamma}$ at the asymmetric point defined by

$$\alpha = \gamma \neq \beta, \qquad p_1 = -p_3,\ p_2 = 0,\ p_3 = p_\beta, \tag{8}$$

then it can be written as

$$\Gamma^{(3)}_{cont\ \alpha\beta\alpha}(-p_\beta, 0, p_\beta) = 2\ \mathcal{A}(p^2)\ p_\beta. \tag{9}$$

The above expression is proportional to the continuum tree-level vertex evaluated at (8).

One can show that the leading term of the 1-loop lattice calculation of the 3-gluon vertex for the same kinematics is indeed consistent with (9) [11], thus when calculating the lattice vertex function $\Gamma^{(3)}_{lat}$ at the point (8) one can set (neglecting terms of higher order in the external momentum):

$$\Gamma^{(3)}_{lat\ \alpha\beta\alpha}(-p_\beta, 0, p_\beta) = 2\ \mathcal{A}_{lat}(p^2, a)\ p_\beta, \tag{10}$$

where $a$ is the lattice spacing. Then, when $a \to 0$, one has

$$\mathcal{A}_{lat}(p^2, a)|_{p^2=q^2} = Z_g^{-1}(a^2q^2)\ g_o(a), \tag{11}$$

for a generic momentum $q^2$ ($g_o(a)$ is the bare lattice coupling constant). Then we obtain the relation

$$Z_g^{-1}(a^2q^2)\ g_o(a) = \frac{1}{2p_\beta}\ \Gamma^{(3)}_{lat\ \alpha\beta\alpha}(-p_\beta, 0, p_\beta)|_{p^2=q^2}. \tag{12}$$

Analogously, one defines the gluon wavefunction renormalization constant $Z_A$ from the relation

$$G^{(2)}_{lat\ \mu\nu}(p^2)|_{p^2=q^2} = T_{\mu\nu}(q)\ Z_A(a^2q^2)\ \frac{1}{q^2}, \tag{13}$$

with $T_{\mu\nu}(q)$ being the projector on transverse fields (we will work in the Landau gauge).

Finally, following Ref. [11], we define the renormalized, "running" coupling as

$$g_R(q^2) = Z_A^{3/2}(a^2q^2)\ Z_g^{-1}(a^2q^2)\ g_o(a). \tag{14}$$

At this point, a standard 1-loop calculation may be performed to relate the coupling constant as defined above to more popular continuum schemes, like for example $\overline{MS}$, etc.

One drawback of the above procedure is that one needs an independent determination of the value of the inverse lattice step size $\frac{1}{a}$ in units of energy. It has been recently pointed out that a very precise determination of such a parameter can now be obtained from the analysis of the charmonium spectrum on the lattice [2]. Also, the determination of $\frac{1}{a}$ from string tension measurements has now reached a high level of accuracy [12].

On the other hand, a very attractive feature of the above method is that in principle it can be applied to the unquenched case without any modification.

It is important to emphasize that in the above procedure one is assuming that lattice correlation functions in momentum space can be effectively parametrized according to continuum formulas. In this respect, provided that $\beta$ is sufficiently large, it is crucial to study the role of finite lattice spacing and finite volume artifacts. The former ones should be under control as long as the momenta that we inject lie well below the value of the inverse lattice spacing. In other words, when injecting the momentum $k_x = \frac{2\,\pi\,n}{L_x\;a}$ along the lattice $x$ direction, with $n$ an integer, we require $k_x$ to satisfy the condition $k_x \ll \frac{1}{a}$. This gives

$$n \ll \frac{L_x}{2\,\pi}. \tag{15}$$

Finite volume effects are potentially more dangerous since the kinematics we have chosen for the definition of the vertex requires the evaluation of the lattice Fourier transform at zero momentum. In this exploratory work we will focus on the investigation of finite volume effects.

As far as the actual calculation of the coupling constant is concerned, one is interested in evaluating it at momentum scales well above 1 GeV, where a comparison to the perturbative result may be attempted and confinement effects can be neglected. In particular, evaluating the coupling at $q = m_b \approx 5$ GeV would allow a comparison to the experimental result [13].

On the lattices which are presently available to us, one cannot inject momenta above $\approx 1.5$ GeV and still assume that finite lattice spacing artifacts are negligible. For this reason, the results presented here are not suitable for the extraction of the coupling constant. Nonetheless, they provide evidence that the 3-gluon vertex function can be effectively measured so that our method for the evaluation of the running coupling may be applicable at larger $\beta$.

## II. THE GLUON PROPAGATOR

The first step in our program is the evaluation of $G^{(2)}_{lat}(p^2)$. The lattice gluon field can be defined as [5]:

$$A_{lat\ \mu} \equiv \frac{U_\mu - U_\mu^\dagger}{2iag_o} - \frac{1}{3} Tr \left( \frac{U_\mu - U_\mu^\dagger}{2iag_o} \right) . \qquad (16)$$

For the purpose of defining the amputated vertex function, we consider a set of 25 configurations on a $16^3 \times 40$ lattice at $\beta = 6.0$ (the same considered in [8]) and 33 configurations on a $24^3 \times 40$ at the same value of $\beta$ (an enlargement of a set considered in Ref. [8], where a better gauge-fixing accuracy has been achieved).

By comparing the results from such sets we hope to gain insight into the role of finite volume effects.

All the calculations are performed after gauge-fixing to the so-called minimal Landau gauge [14,15] (see Ref. [8] for a short review), implemented through the iterative minimization of

$$H_U[g] \equiv -\frac{1}{V} \sum_{n,\mu} \ Re\ Tr\ \left( U^g_\mu(n) + U^{g\dagger}_\mu(n - \hat{\mu}) \right) , \qquad (17)$$

where $V$ is the lattice volume and $U^g$ indicates the gauge-transformed link $U^g_\mu(n) \equiv g(n) U_\mu(n) g^\dagger(n + \hat{\mu})$. The typical precision that we reach for such a minimization is $\approx 10^{-5}$. In Fig. 1 we show our results for the momentum space gluon propagator on the $16^3 \times 40$ lattice.

## III. THE 3-GLUON VERTEX

After calculating the propagator, we proceed to the evaluation of the complete 3-point function $G^{(3)}_{lat\ \alpha\beta\alpha}(-p_\beta, 0, p_\beta)$. As our main purpose is to test the quality of the signal and the occurrence of finite volume effects, we proceed as follows. First we set $\alpha = i$, $i = 1, \ldots, 3$, $\beta = 4$, injecting momentum in the longer (time) lattice direction, and we call $p_t$ the momentum. In this case we can safely inject a few units of momentum while satisfying the requirement given by Eq. (15), but finite volume effects are expected to be most severe as the zero-momentum Fourier transform is performed along all the (shorter) spatial axes. To improve statistics one observes that from Eq. (7) it follows that at the asymmetric point:

$$\Gamma^{(3)}_{cont\ 1\ 4\ 1}(-p_t, 0, p_t) = \Gamma^{(3)}_{cont\ 2\ 4\ 2}(-p_t, 0, p_t)$$
$$= \Gamma^{(3)}_{cont\ 3\ 4\ 3}(-p_t, 0, p_t). \tag{18}$$

Obviously an analogous relation holds for $G^{(3)}$, so that we can average $G^{(3)}$ and $\Gamma^{(3)}$ over the spatial Lorentz components.

After this calculation, we change the kinematics. We define an index $\mu^i$ which runs on all lattice directions but the $i$ one. Then we set $\alpha = \mu^i$, $\beta = i$, $i = 1, \ldots, 3$ and we inject a momentum $p_i$ in the $i$ direction. In this case not only we expect finite volume effects to be less severe, but we also have better statistics, as the symmetries of (7) allow to average over 9 distinct vertex functions. Finally, we compare the results from the 2 sets of lattices.

Consider first the $16^3 \times 40$ lattice. The complete 3-point function $G^{(3)}_{lat}(-p, 0, p)$ vs. $p$ is shown in Fig. 2 for both the kinematical arrangements described above. The points corresponding to space and time momenta are plotted with different symbols. It appears that for large values of the momentum $p$, $G^{(3)}(-p, 0, p)$ gets damped, due to the effect of the propagators lying on the external legs. We observe that the data points are roughly consistent with an odd function of the injected momentum, as expected. In this respect, the points obtained from spatial momentum, besides having smaller error bars than the others, turn out to be antisymmetric to a better accuracy.

In order to define the amputated vertex, we multiply the complete 3-point function shown above by the inverse propagators, according to Eq. (3). The resulting function $\Gamma^{(3)}_{lat}(-p, 0, p)$ is shown in Fig. 3, where the error is a jackknife one.

As the vertex must be an odd function of the momentum, we can further improve statistics by antisymmetrizing. The results are shown in Fig. 4.

Turning now to the bigger lattice, i.e. the $24^3 \times 40$ one, we show in Fig. 5 the amputated vertex function after antisymmetrization.

By comparing the results in Figs. 4 and 5 it appears that, at the level of precision that we can presently reach, finite volume effects are not detected, as the data points corresponding to the injection of momentum in time are consistent between the two lattices.

As expected, the data points corresponding to spatial momenta are the most stable. We collect such points from both lattices in Fig. 6.

We observe that, in spite of the better statistics, the results from the bigger lattice are noisier. This may be related to an insufficient gauge-fixing accuracy. To investigate this issue we have performed some tests on a single configuration, studying the stability of our results vs. additional gauge-fixing. Such tests seem to indicate that the gauge-fixing accuracy was sufficient for our purposes, although this point deserves further investigation.

In this connection, it is worth remarking that the accuracy of the numerical gauge-fixing is a very important parameter in our calculations. In fact, it turns out that for the purpose of evaluating gluon 3-point functions one needs a higher accuracy than when calculating the propagator. To illustrate this point, in Figs. 7 and 8 we plot the propagator and the complete 3-point function respectively, as evaluated on a single configuration of the $24^3 \times 40$ lattice, for different levels of gauge-fixing precision.

From Fig. 7 one can see that the propagator from 600 to 1400 gauge-fixing sweeps is virtually unchanged. On the other hand, Fig. 8 shows that for the 3-point function 600 gauge-fixing sweeps are definitely not enough.

## IV. CONCLUSIONS

We have discussed a method to measure on the lattice the 3-gluon vertex function in momentum space and to extract from it the running coupling constant. We have given numerical results at $\beta = 6.0$ for the vertex function. Such results are characterized by a good signal-to-noise ratio and show the symmetries expected. As a consequence, it appears that the operation of "amputating" at the non-perturbative level the lattice 3-point function in momentum space is feasible and leads to sensible numerical results, even with limited statistics.

By comparing results for two different lattice sizes it appears that finite volume effects do not play a significant role at our current level of precision.

As we already remarked, our results are not yet suitable for a realistic determination of the QCD running coupling constant, both because of the level of statistical noise and because of the momentum range under consideration. Nonetheless, they suggest that such a task may be feasible in a high statistics calculation at weaker couplings.

## ACKNOWLEDGMENTS

The author acknowledges support from SERC under grant GR/J 21347 and partial support under USDOE contract number DE-AC02-76CH00016. The computing was done at the National Energy Research Supercomputer Center in part under the "Grand Challenge" program and at the San Diego Supercomputer Center. The author warmly thanks C. Bernard and A. Soni for continuous support and many illuminating discussions, and D. Henty and G. Martinelli for interesting discussions.

## FIGURES

FIG. 1. $G^{(2)}_{lat}(p^2)$ vs. $p$ in GeV on the $16^3 \times 40$ lattice at $\beta = 6.0$. We assume $a^{-1} = 2.1$ GeV.

FIG. 2. $G^{(3)}_{lat}\,(-p, 0, p)$ vs. $p$ in GeV on the $16^3 \times 40$ lattice. Crosses and diamonds correspond to the injection of momentum in space and time directions, respectively.

FIG. 3. $\Gamma^{(3)}_{lat}\,(-p, 0, p)$ vs. $p$ in Gev on the $16^3 \times 40$ lattice. Crosses and diamonds correspond to the injection of momentum in space and time directions, respectively.

FIG. 4. Same as in Fig. 3, after imposing antisymmetry.

FIG. 5. Same as in Fig. 4, for the $24^3 \times 40$ lattice.

FIG. 6. $\Gamma^{(3)}_{lat}\,(-p, 0, p)$ vs. $p$ in Gev for spatial momenta. Crosses and diamonds correspond to the $24^3 \times 40$ and the $16^3 \times 40$ lattices, respectively.

FIG. 7. $G^{(2)}_{lat}(p^2)$ vs. $p$ in GeV calculated on a single configuration of the $24^3 \times 40$ lattice with different gauge-fixing accuracies.

FIG. 8. $G^{(3)}_{lat}\,(-p, 0, p)$ vs. $p$ in GeV calculated on a single configuration of the $24^3 \times 40$ lattice with different gauge-fixing accuracies.

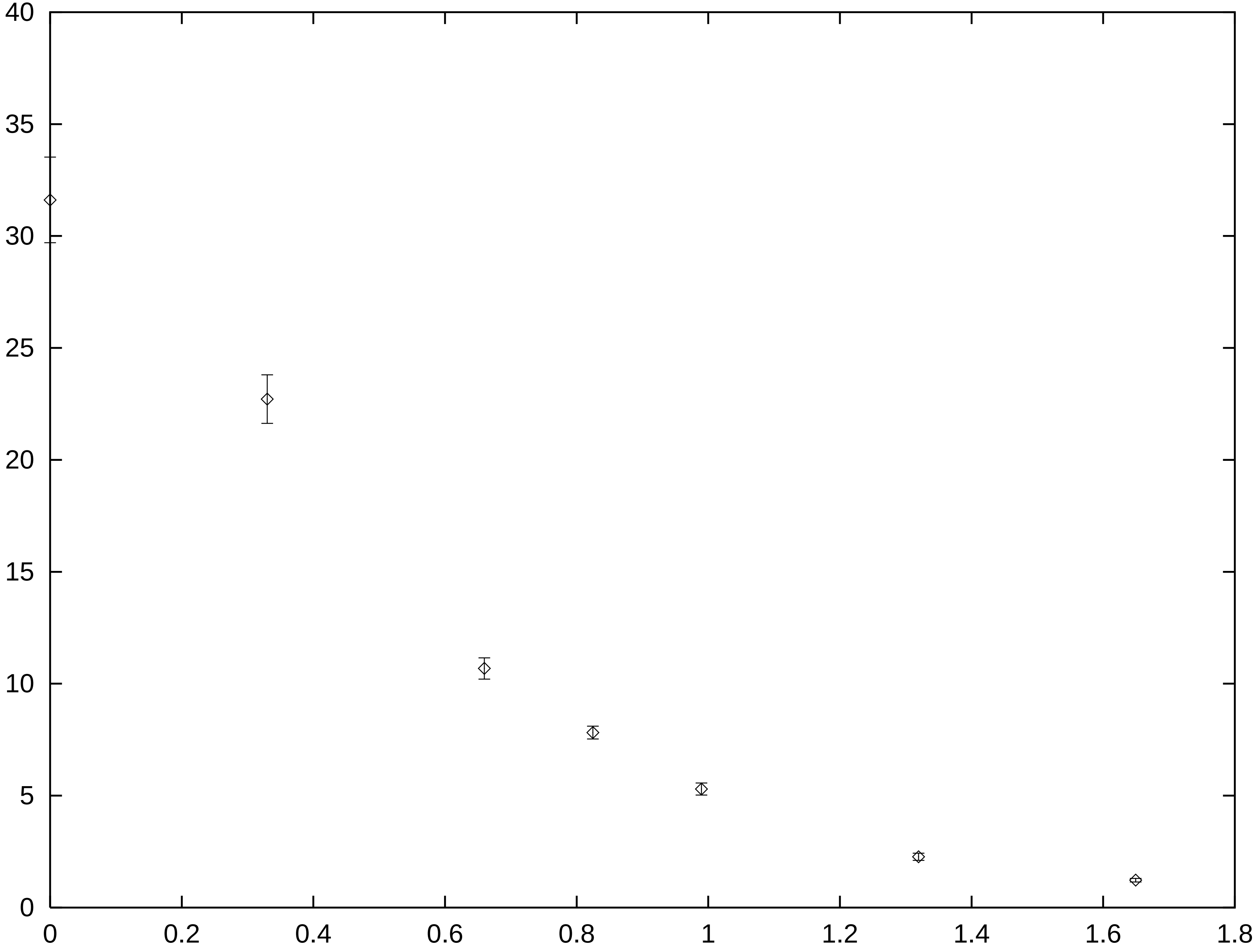

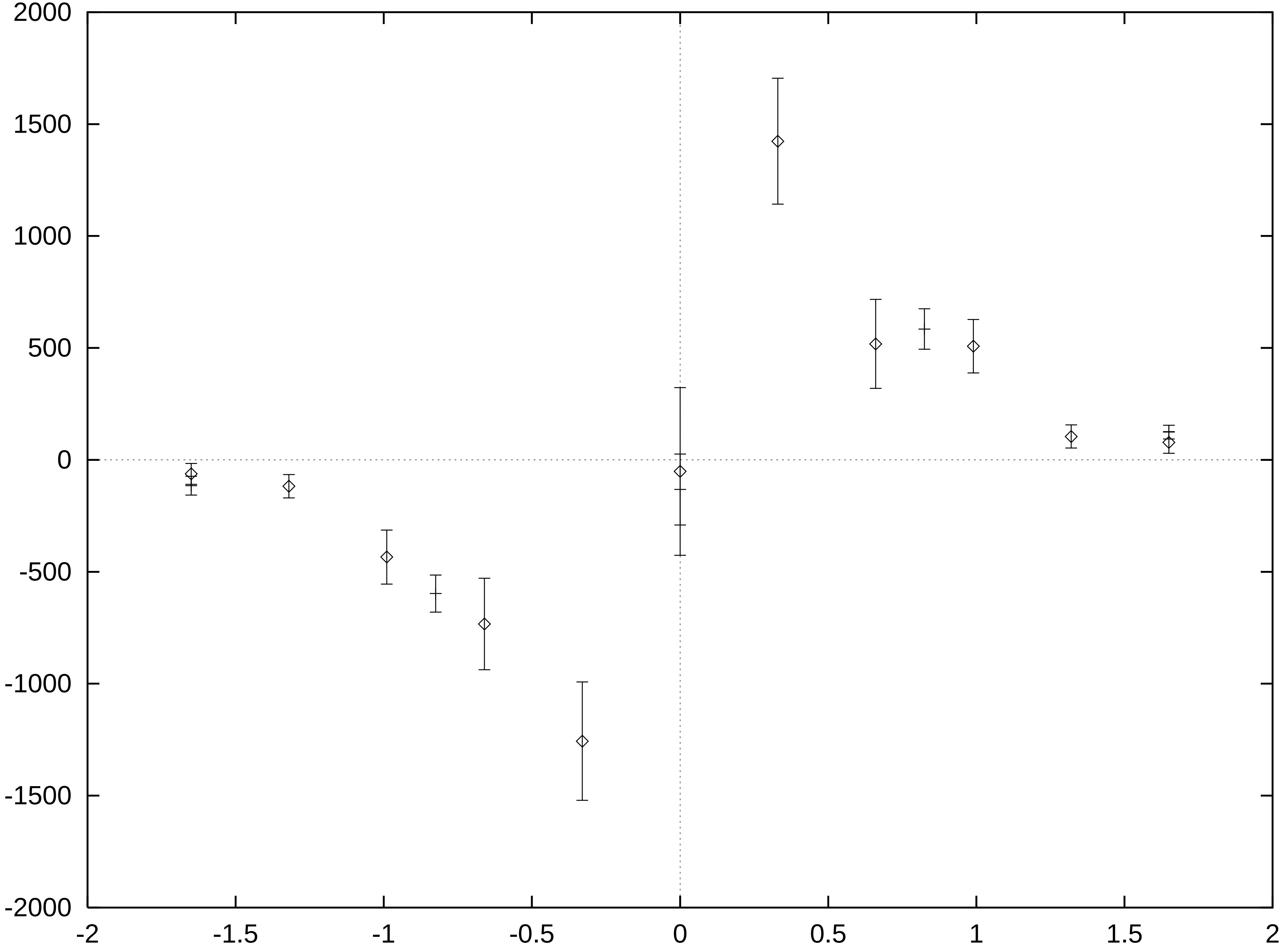

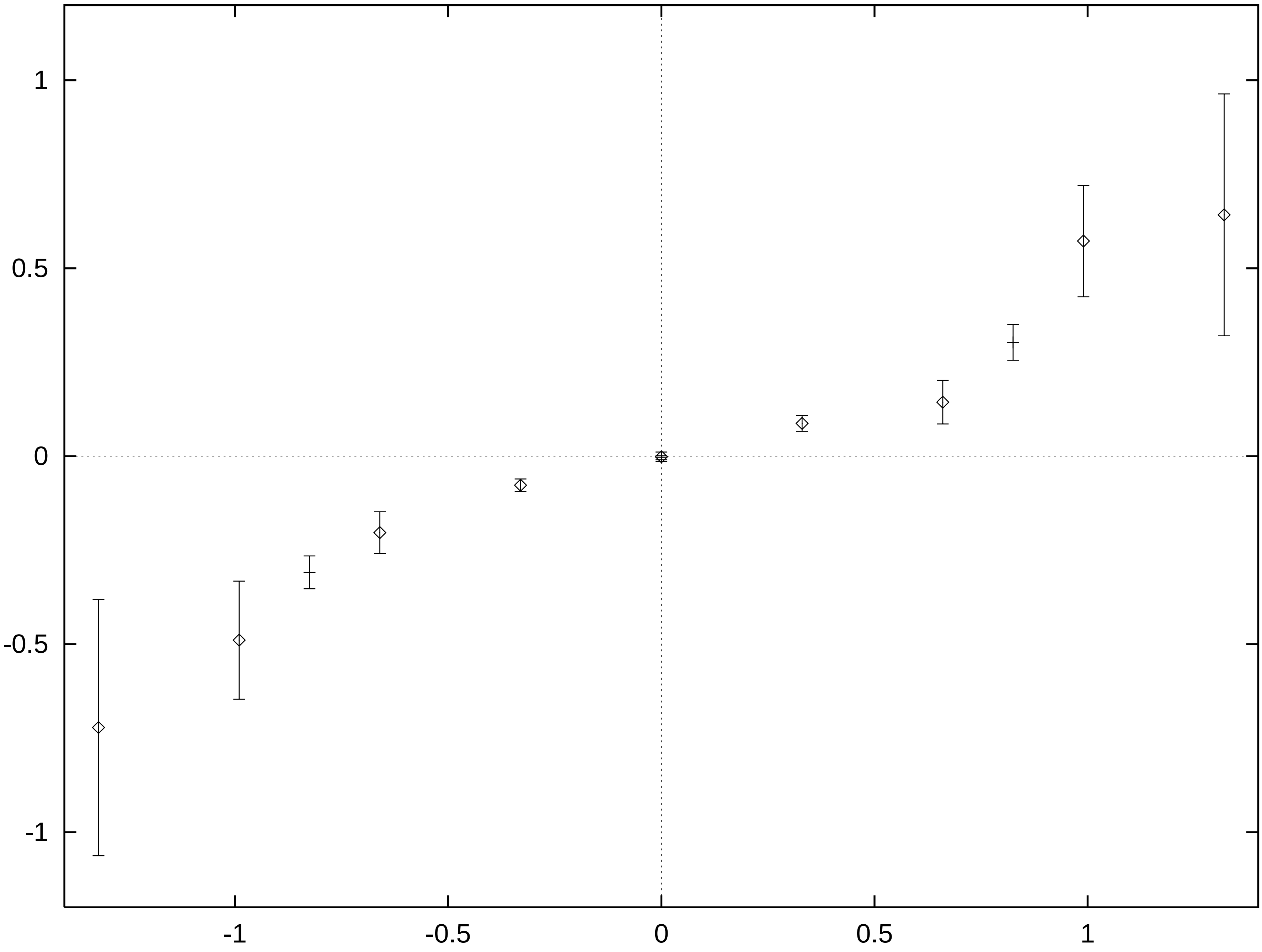

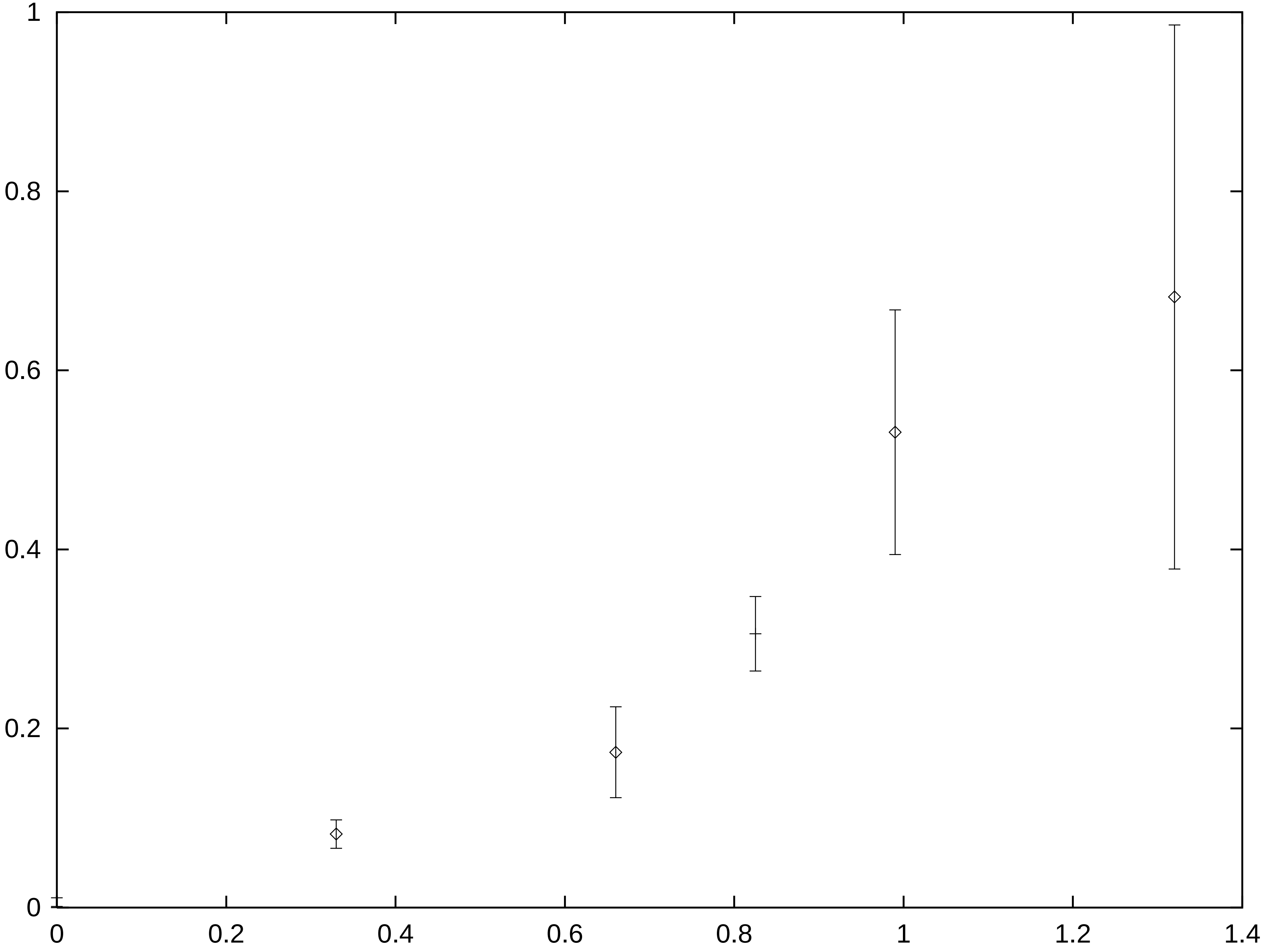

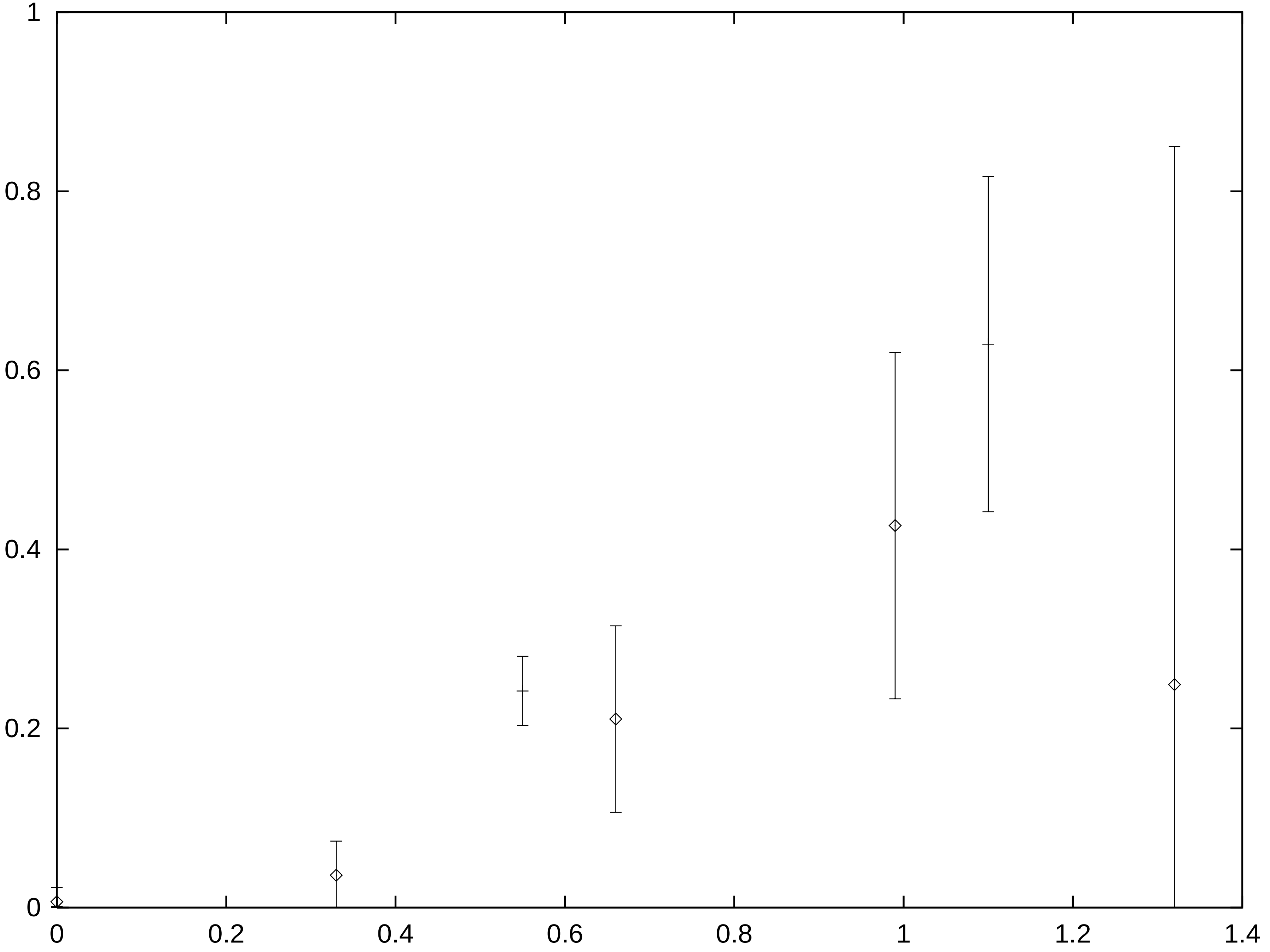

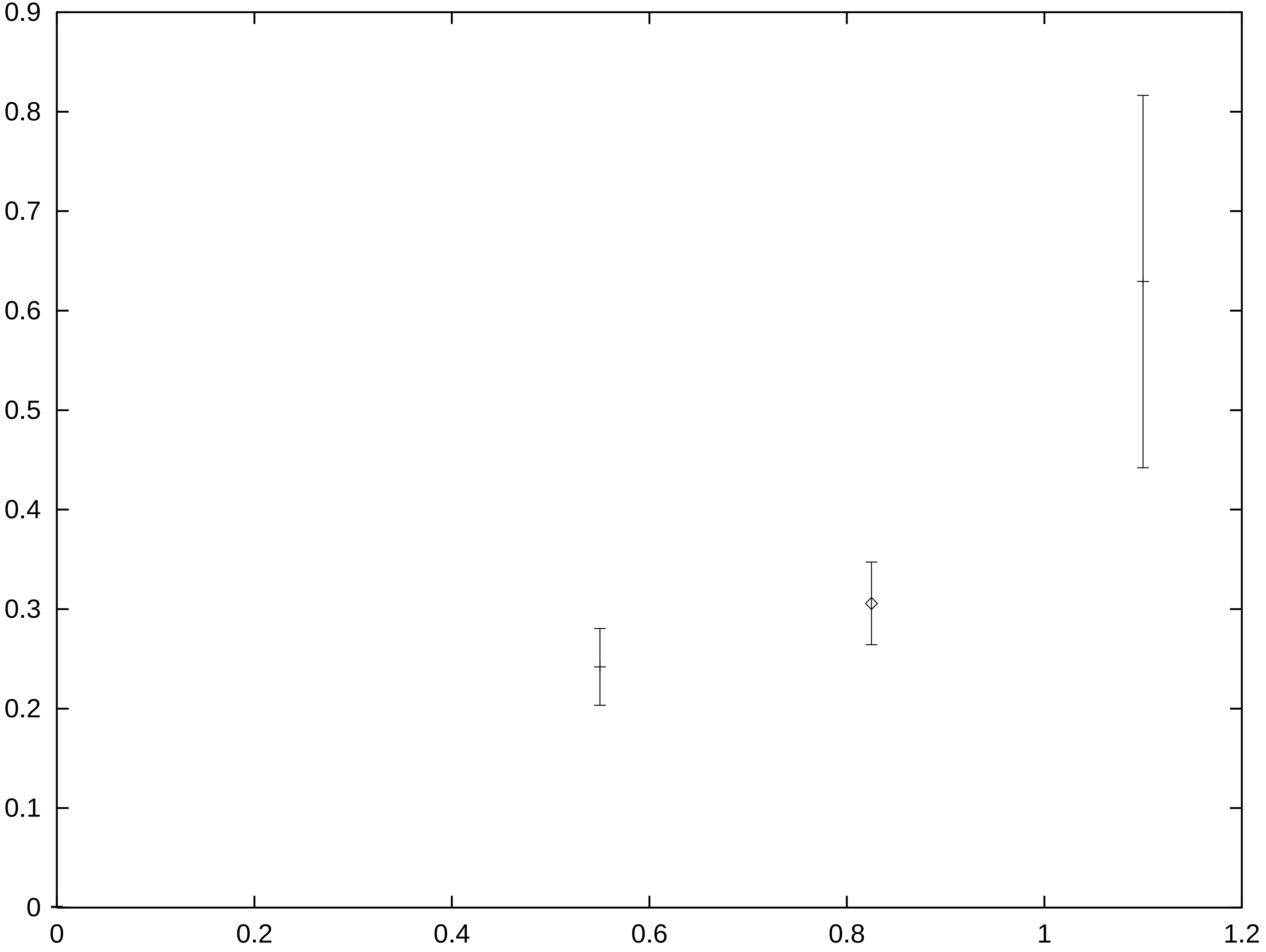

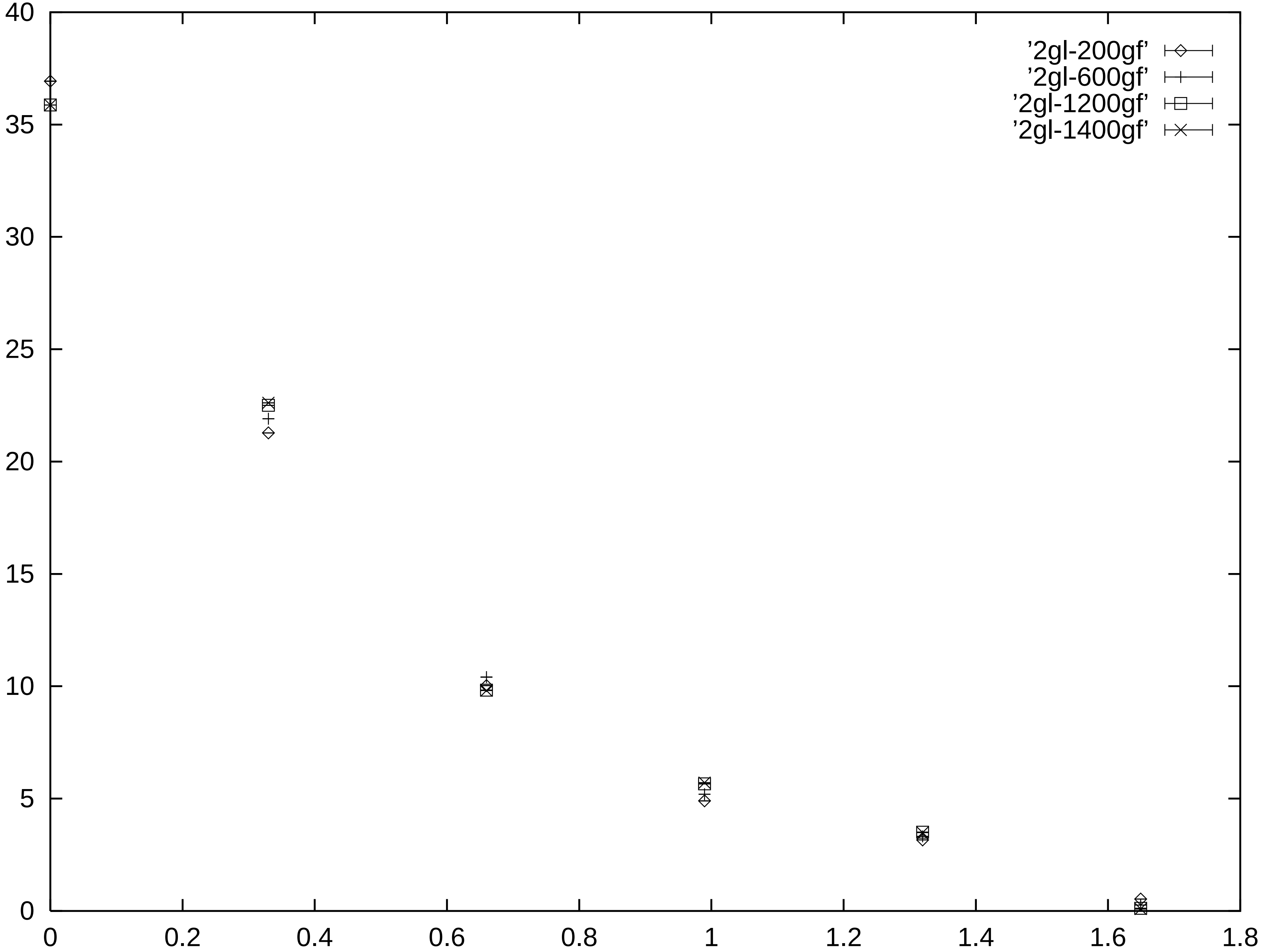

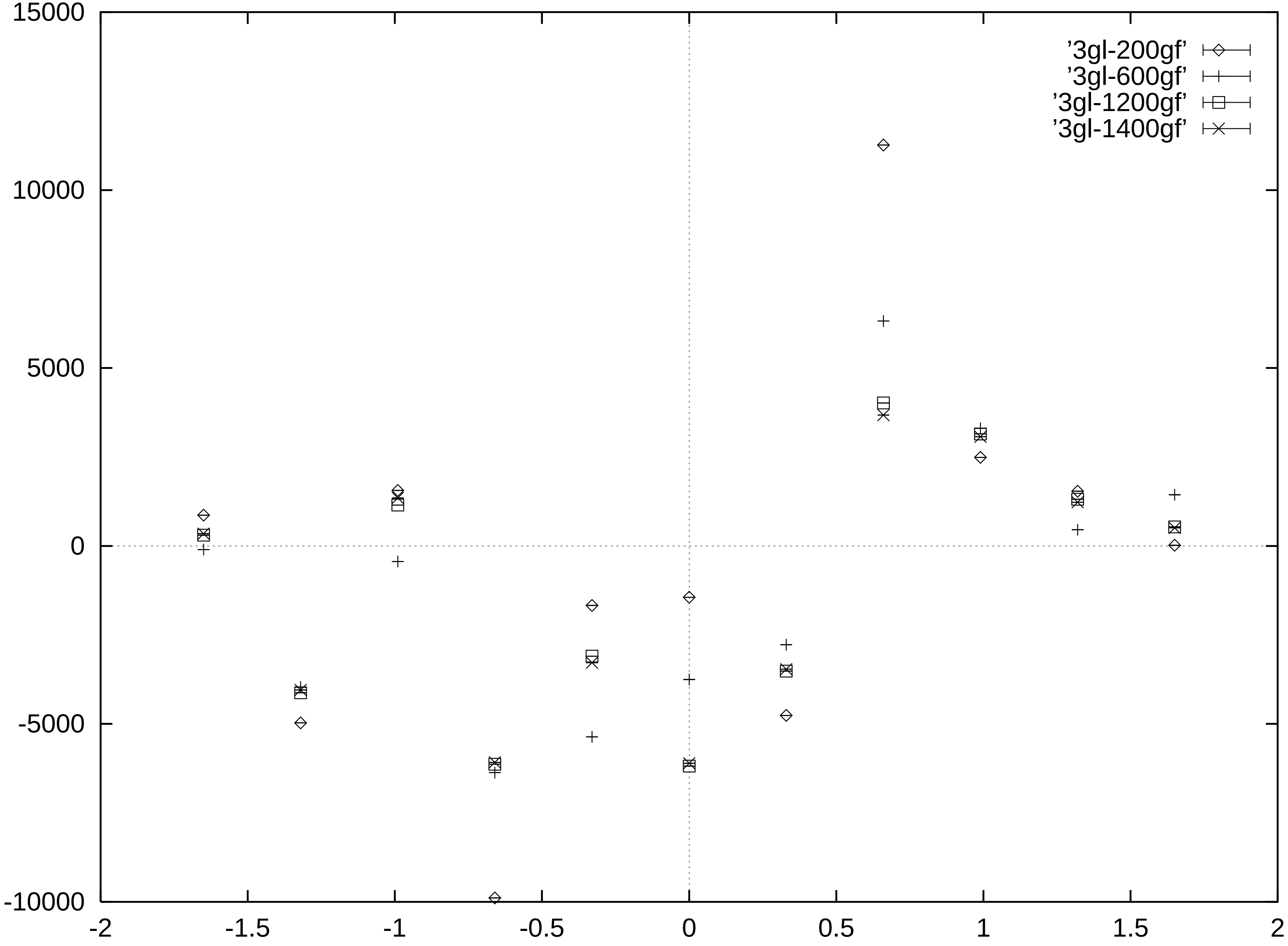